# Monitoring in IOT enabled devices

Udit Gupta

Information Networking Institute

Carnegie Mellon University, Pittsburgh – Pennsylvania, USA

uditg@andrew.cmu.edu

**ABSTRACT:**

As network size continues to grow exponentially, there has been a proportionate increase in the number of nodes in the corresponding network. With the advent of Internet of things (IOT), it is assumed that many more devices will be connected to the existing network infrastructure. As a result, monitoring is expected to get more complex for administrators as networks tend to become more heterogeneous. Moreover, the addressing for IOTs would be more complex given the scale at which devices will be added to the network and hence monitoring is bound to become an uphill task due to management of larger range of addresses. This paper will throw light on what kind of monitoring mechanisms can be deployed in internet of things (IOTs) and their overall effectiveness.

1. **INTRODUCTION:**

Monitoring [3, 4, 5 and 6] of systems has always been of prime importance for organizations since the advent of internet in 1960's. The abrupt increase in the number of devices ensured that manually monitoring every device connected to the network would become infeasible. As internet expanded to more number of devices which includes wireless devices, more robust monitoring systems were designed which were simple to use yet powerful enough to detect failures. As the wave of virtualization swept the technology world in early 2000's, monitoring systems had to be evolved in order to monitor virtual machines. As cloud computing [19 and 20] came into existence over the past decade, monitoring systems had to be customized according to cloud technology. As world is looking towards IOTs, it becomes paramount to design monitoring systems which will be compatible with different kinds of devices and at the same time detect intrusions and failures efficiently.

The term 'Internet of things' or IOTs [7, 8 and 9] refers to the networking amongst everyday objects or with a centralized system. More colloquially, it is the process of how various 'things' which are connected to the internet interact with each other. Ever since the idea of first internet based coke machine was conceived at Carnegie Mellon University in 1982, the idea about extending internet to entities turned into a reality. This can be observed in the form of RFID [10, 11 and 12] technology which provides a unique identity to objects. Apart from this, iPad controlled lights has gained significant importance in organizations and homes. To add to this list, there has been a gradual progress in the research on smart cars and smart homes [13]. Furthermore, in order to make IOT a reality a lot of research has been going on in domain of security issues [1] pertaining to IOT and operating systems [2] which will be compatible with IOT. It's important to have monitoring systems in place which would ensure that these devices are monitored periodically and intimates user accordingly.

There are three primary reasons for designing monitoring systems: intrusion detections [14, 15, 21, 24, 25 and 26], notifications about critical parameters reaching threshold and alerts pertaining to failure of the corresponding machine. Therefore, it's important that whatever monitoring systems are deployed onto the architecture they must analyze all three parameters named above and must notify user accordingly. Moreover, if the monitoring system can identify whether encrypted [22, 23 and 24] connection exists between two nodes then it'll be of added advantage to the network as a whole. In this paper, we will discuss two monitoring agents (Big Brother and Zenoss) and how they can be utilized in monitoring IOTs.

## 2. MONITORING SYSTEMS FOR IOTs

Over the past decade, a number of monitoring systems were designed for effective network administration. Monitoring data will be present in almost all the nodes in the network but an efficient system will be able to collect all this data and present it to administrator in a comprehensive manner. Some of the monitoring systems which could meet this criterion are: Nagios, Big Brother, Ntop, Zenoss and Multi Router Traffic Grapher (MRTG). Big Brother, Zenoss and Nagios are mostly used as part of enterprise level monitoring solution since they could monitor all aspects of the network. On the other hand, the functionality of Ntop and MRTG are somewhat limited since they were primarily designed to monitor IP traffic. Our prime focus henceforth will be on Big Brother and Zenoss and how they could be deployed in IOTs.

### a. USING BIG BROTHER (BB) IN IOT

Big Brother (also called BB) is the first monitoring system to use web as its user interface. This allowed non-technical users to see the holistic picture and gain an understanding of network in a simplified manner. BB uses a client-server model where client will send status information to monitoring server after certain time intervals. BB can be effective tool to monitor IOTs. Consider the example of smart home. Consider an example of smart home which will have smart refrigerator, smart washing machine and smart AC. Let's consider the definitions of each entity described. Smart refrigerator will adjust its temperature automatically based on the weather so that electricity consumption can be optimized. Smart washing machine will automatically set its washing mode based on the cloth material and finally smart AC will again be able to adjust its temperature according to climate outdoors. Each of these smart devices will operate on sensors or actuators or both. Hence it is important to maintain a server and a UI, which will present required information to user on his/her mobile/tablet. Also a tabular form can be maintained as shown in table 1 where for sensors/actuators 'working' condition is highlighted in green while 'not working' condition is highlighted in red. For memory/cpu-utilization three colors are used to indicate the value above or below threshold. This table is just an excerpt of a larger table which will have many more smart things and corresponding parameters to be analyzed.

### b. USING ZENOSS IN IOT

Zenoss is an open source platform based on SNMP [16 and 17] protocol and it monitors networks, servers, applications and services. The biggest advantage of this monitoring system is that it is an open architecture and hence users can customize it according to their own needs. As a result, it is widely used in enterprise solutions. The functionality of this monitoring system can be extended to IOTs as well. Given that SNMP is the underlying protocol, the notification mechanism can be either polling or traps. Polling would be a better option for most IOT based devices since metrics can be stored and as a result historical analysis can be performed. Also configuring notifications and alarms based on active polling is much easier as compared to traps. Consider again the example of smart home described previously: network server can easily identify via polling the status of smart things. Metrics will be presented to user in the form of UI and it will also include data in graphical form.

### 3. COMPARISON BETWEEN ZENOSS AND BB IN IOT

Both BB and Zenoss have been extensively used by enterprises in the past. Given their wide range of functionality and ease of deployment they still continue to be a part of most organizations monitoring framework. As a result of their ability to provide comprehensive overview of the network, both can be deployed for monitoring IOT enabled devices. Both can work in heterogeneous environment where apart from virtual or physical machines other network devices like routers, switches can also be monitored.

But after due consideration of the features being offered by these two monitoring agents, Zenoss is found to be slightly better than BB due to multiple reasons. Zenoss is found to support IPv6 protocol which is bound to gain paramount importance in case of IOTs due to range of IP addresses being offered by it. As mentioned previously, there will be an exponential increase in the number of IOT enabled devices in the near future and as a result identifying each device uniquely will be of prime importance. Therefore, it's essential to have a protocol in place which will ensure that numbers of unique tags are not depleted while identifying IOT devices. IPv6 serves this purpose quite well. With Zenoss being compatible with IPv6, its execution in IOT will be less complex.

Secondly, in Zenoss there is provision for both: polling as well as trap, both of which will be significant as far as IOT is concerned. Let us consider an example: if a smart car is being monitored then it is important to gain a periodic update as to whether it is working properly or not. Thus, implementing polling becomes inevitable. On the other hand if something needs to be communicated on an urgent basis like calling a fire brigade by a smart device then implementing trap is the only feasible option. With SNMP offering both these features and given that Zenoss is based on SNMP, it can safely be assumed to be ideal for IOT.

Finally, it'll easier to manage IOT devices using Zenoss due to the fact that an MIB [18] file can be created for each managed object. MIB (management information base) file is a formal description of a set of network objects that can be managed using the SNMP. A sample MIB file for smart home described above has been mentioned below:

```
SMART_HOME-MIB DEFINITIONS: = BEGIN
IMPORTS
     enterprises
          FROM RFC1155-SMI
     OBJECT-TYPE
          FROM RFC-1212
     DisplayString
          FROM RFC-1213;
smartAC OBJECT-TYPE
   SYNTAX  DisplayString
   ACCESS  read-only
   STATUS  mandatory
   DESCRIPTION
            "The description of the smart AC"
smartRefrigerator OBJECT-TYPE
   SYNTAX  DisplayString
   ACCESS  read-only
   STATUS  mandatory
   DESCRIPTION
            "The description of the smart refrigerator"
smartWashingMachine OBJECT-TYPE
   SYNTAX  DisplayString
   ACCESS  read-only
   STATUS  mandatory
   DESCRIPTION
            "The description of the smart washing machine"
```

Table 1

|  | Sensor | Actuator | Memory | CPU utilization |
|---|---|---|---|---|
| Smart AC | Working | Working | Below threshold | Below Threshold |
| Smart Refrigerator | Not working | Working | Threshold | Below Threshold |
| Smart washing machine | Working | Not working | Above Threshold | Above Threshold |

## 4. CONCLUSION

While Internet of things (IOT) is yet to mark its arrival in a substantial way in technology world, it becomes paramount to address monitoring perspective beforehand. Some of them were highlighted in this paper and a brief attempt was to enable the reader to understand the context in which two of the existing monitoring solutions can be deployed for IOT enabled devices. Although only two monitoring systems were presented in this paper which could be deployed in IOT, there is no denial that other monitoring systems in the market can also be implemented as part of IOT enabled devices. But there is no denying the fact that Big Brother and Zenoss can monitor wide range of applications which includes processes, events and even logs. A comparison was presented between these two monitoring systems and then an inference was drawn based on certain observations about which would be better in case of IOT.